# Proposal for nanoscale cascaded plasmonic majority gates for non-Boolean computation

Sourav Dutta[1*], Odysseas Zografos[2,3], Surya Gurunarayanan[2,3], Iuliana Radu[2], Bart Soree[2,3,4], Francky Catthoor[2], Azad Naeemi[1]

[1]School of Electrical and Computer Engineering, Georgia Institute of Technology, Atlanta, Georgia 30332, USA
[2]imec, B-3001 Leuven, Belgium
[3]KU Leuven, ESAT, B-3001 Leuven, Belgium
[4]Universiteit Antwerpen, Physics Department CMT, B-2020 Antwerpen, Belgium

*Correspondence to sdutta38@gatech.edu

**Surface-plasmon-polariton waves propagating at the interface between a metal and a dielectric, hold the key to future high-bandwidth, dense on-chip integrated logic circuits overcoming the diffraction limitation of photonics. While recent advances in plasmonic logic have witnessed the demonstration of basic and universal logic gates, these CMOS oriented digital logic gates cannot fully utilize the expressive power of this novel technology. Here, we aim at unraveling the true potential of plasmonics by exploiting an enhanced native functionality - the majority voter. Contrary to the state-of-the-art plasmonic logic devices, we use the phase of the wave instead of the intensity as the state or computational variable. We propose and demonstrate, via numerical simulations, a comprehensive scheme for building a nanoscale cascadable plasmonic majority logic gate along with a novel referencing scheme that can directly translate the information encoded in the amplitude and phase of the wave into electric field intensity at the output. Our MIM-based 3-input majority gate displays a highly improved overall area of only 0.636 µm$^2$ for a single-stage compared with previous works on plasmonic logic. The proposed device demonstrates non-Boolean computational capability and can find direct utility in highly parallel real-time signal processing applications like pattern recognition.**

For more than four decades, Moore's law has been the driving force for the semiconductor industry. However, as CMOS circuits approach the scaling limitations due to fundamental physical constraints, this scaling trend is eventually going to end at some point in the future [1,2]. Photonic devices and circuits are a promising alternative due to their high speed and low propagation loss [3]. However, the diffraction limit of light proves to be an obstacle for realizing nanoscale photonic devices as the dimensions approach the wavelength of light in the material. Surface plasmon polaritons (SPP) - electromagnetic waves propagating at the interface between a metal and a dielectric - can circumvent this problem by localizing the electromagnetic energy in dimensions much smaller than the diffraction limit [4,5]. Recent advances in the field of plasmonics have witnessed the development of innovative waveguiding schemes [6-12] and devices [13-20]. A complete set of fundamental logic gates (basic and universal) have been realized by exploiting phase-dependent interference of SPP waves [18-20]. However, these CMOS based digital logic gate designs cannot exploit an important feature of plasmonic logic and wave computing, namely, the ability to execute majority voting efficiently. Deviating from the traditional path of CMOS oriented chip design, new logic abstractions and synthesis techniques are being developed for the emerging nano-devices that are capable of reproducing the same CMOS logic circuitry with lower footprint and higher performance with only two building blocks – inverter and majority logic gate [21]. While recent works in the analogous field of spin wave based computing [22] have seen the proposal for a spin wave majority logic gate [23], a wave computing paradigm based on majority gates has not yet been implemented with plasmons.

In this work, we aim at highlighting the enhanced native functionality of plasmonics based logic - the majority gate - utilizing the phase of the SPP wave as the state variable. Using finite-difference-time domain (FDTD) simulations performed in commercially available software Lumerical Solutions [24], we provide a comprehensive scheme for building a nanoscale cascadable plasmonic majority logic gate. We start by characterizing the waveguiding properties of the chosen slot waveguide followed by the design and simulation results of a single stage 3-input plasmonic majority gate. We demonstrate that besides Boolean logic, our proposed plasmonic majority gate is also capable of performing non-Boolean computing due to multi-level output denoting the strength of the majority. We investigate the cascadability of these gates by studying up to 3 stages of cascaded logic. This is sufficient to support interesting arithmetic primitives like adders and multipliers with limited bit-width. Larger arithmetic processor data-paths can then be composed from these primitives, however this lies outside the scope of this paper. We also propose a unique referencing scheme at the output that can directly translate the information encoded in the amplitude and phase of the output SPP wave into the intensity of the output electric field or the output power, thus circumventing the challenging problem of tera-Hertz phase-detection of the SPP waves. Due to high throughput, the proposed scheme can be of use in highly parallel real-time signal processing applications that are arithmetic-heavy with strict timing requirements. A representative example of this, namely a pattern recognition system, has been briefly discussed towards the end of the paper. Note that since the emphasis of this work is on building a nanoscale cascadable plasmonic majority logic, we do not focus on the mechanism or circuits for excitation and final detection of SPP which are the focus of separate works in the literature [25-37].

## Results

**Characterization of Slot Plasmonic Waveguide**

We start by first describing the basic building block of a plasmonic majority gate – the plasmonic waveguide, that acts as a conduit for transmission of information. We choose a slot (metal-insulator-metal MIM) waveguide configuration due to its high field confinement capability (in nanometers) that aids the designing of nanoscale logic devices [12,20,38]. Fig. 1(a) shows the schematic of the slot plasmonic waveguide. Silver (Ag) on top of a silicon dioxide ($SiO_2$) substrate is used for the metallic structure while the dielectric is assumed to have a refractive index of 1.5. The refractive index of the surrounding air is set to 1. We perform finite-difference time-domain (FDTD) simulations in Lumerical Solutions[24] (see method section for simulation details). We choose the height ($h$) of the metal slot as 100 nm throughout our simulations while the width ($w$) of the waveguide is varied between 60, 120 and 180 nm as explained later.

In a MIM geometry, the two identical SPP modes overlap with each other for small dielectric (insulator) layer width, resulting in a so-called gap SPP (G-SPP). The G-SPP mode displays an odd symmetry of the longitudinal electric field component $E_x$ and even symmetry of the transverse field component $E_y$. The electromagnetic fields are predominantly confined within the slot with the evanescent decay $e^{-k_y y}$ inside the metal, where $k_y$ is the transverse component of the wave vector (y-direction). Along the direction of propagation, the electromagnetic fields vary harmonically as $e^{i(k_x x - \omega t)}$, where $k_x$ is the longitudinal component of the wave vector (x-direction). The longitudinal wave vector $k_x$ can be defined in terms of the permittivities of the metal $\epsilon_m$ and dielectric $\epsilon_m$ and the free-space wavelength of light $\lambda_0$. The real part of the SPP wave vector can be further used to define the effective index of the plasmonic mode inside the waveguide as $n_{SPP}^{eff} = \frac{\lambda}{2\pi} Real(k_x)$. Using appropriate boundary conditions for the normal and tangential electric field components and the aforementioned symmetry of the corresponding field component distributions, the G-SPP dispersion relation can be obtained as $\tanh\left(\frac{k_y^d w}{2}\right) = -\frac{\epsilon_d k_y^m}{\epsilon_m k_y^d}$,[39,40] where, $k_y^{d,m} = \sqrt{k_x^2 - \epsilon_{d,m} k_0^2}$ and $k_0 = \frac{2\pi}{\lambda_0}$. Approximate analytical expressions for the G-SPP dispersion relation for sufficiently small or relatively large gap width have been obtained in refs. [14,40]. Fig. 1(b) shows the numerically calculated dispersion relation of the slot waveguide for three different widths (60, 120 and 180 nm).

Since the emphasis of this work is on building a nanoscale cascadable plasmonic majority logic, we do not refer to a particular technique of excitation (see supplementary section S1 for possible ways of exciting on-chip SPP waves) and use the standard mode source in our Lumerical simulations to inject a fundamental guided mode into the plasmonic waveguide (see supplementary section S2 for details about the excitation source). We choose an operating wavelength of $\lambda_0 = 1.55 \, \mu m$ (corresponding to a frequency of 193 THz) for the mode source in Lumerical simulations. The corresponding wavelength dependent complex permittivity of Ag is -116.676+11.6522i [41] while the permittivity of SiO$_2$ used here is 2.08. Note that the wavelength of the propagating SPP mode ($\lambda_{SPP}$) is different from $\lambda_0$ depending on the width of the waveguide as indicated by the dispersion relation in Fig. 1(b). Fig. 1(c) shows the fundamental plasmonic mode (electric field) distribution inside a 60 nm wide metal-slot waveguide at $\lambda_0 = 1.55 \, \mu m$ suggesting deep subwavelength confinement. The maximum electric field intensity is located at the metal-dielectric interface around the two lower vertexes, consistent with the results of Pan *et. al.* [20].

The development of plasmonic logic is hindered due to the presence of dissipative losses. Dissipative losses arise due to ohmic losses encountered by surface plasmons propagating along the interface of metal and dielectric [42]. The propagation length of surface plasmons depends on the properties of the waveguide – material, geometry, and mode profile of the propagating surface plasmon polariton, and is calculated as the distance over which the propagating power decays up to $1/e$: $L_p = \frac{1}{2\text{Im}[k_x]}$. The evanescent decay length inside the metal $\delta = \frac{1}{\text{Im}[k_y^m]}$ is defined as the distance over which the field decays up to $1/e$ in the transverse direction [43]. Both the propagation length ($L_p$) and confinement ($\delta$) show strong frequency dependence. A higher frequency provides better confinement. However, this would imply an increase in the damping of electrons and energy dissipation in the metal, causing a decrease of the propagation length. Thus, there exists a trade-off between $L_p$ and $\delta$. To quantify this trade-off, next we look at the individual figures of merits (FOM). The figures of merit for propagation and confinement are given by $FOM_{prop} = \frac{L_p}{\lambda_{spp}}$ and $FOM_{conf} = \frac{\lambda_0}{\delta}$ respectively. Note that there are various other definitions of plasmonic mode confinement involving the spatial extent of the energy [38,44]. However, all of them are largely determined by the same exponentially decaying field outside the waveguide, so we believe that our definition of $\delta$ is sufficiently representative. Fig. 1(d) shows the two dimensional figure of merit (FOM) graph [45] for our exploration space as a function of the excitation wavelength $\lambda_0$ of the mode source in Lumerical simulations for the three different widths of the waveguide. The chosen wavelength of $\lambda_0 = 1.55\ \mu m$ illustrates a good trade-off between the propagation and the confinement. While the MIM waveguides feature lower propagation lengths compared to insulator-metal-insulator (IMI) waveguides, the higher confinement results in a tighter pitch (center-to-center distance between the adjacent waveguides) while maintaining a low crosstalk noise. This is critical for signal integrity. The crosstalk noise, defined as the coupling or overlap of modes between the adjacent waveguides resulting in a transfer of power from one waveguide to another, is measured in terms of the coupling length $L_c = \frac{\lambda_0}{\pi \Delta n} sin^{-1}\left(\sqrt{\frac{P_V}{P_A}}\right)$ where $P_A$ and $P_V$ are the powers in the active and victim waveguides respectively and $\Delta n$ is the index difference between the two coupled modes. For the worst case scenario, we calculate the length for 100% coupling of power as $L_c = \frac{\lambda_0}{2\Delta n}$. Fig. 1(e) shows the coupling length $L_c$ as a function of the waveguide pitch $p$ for three different widths of the waveguide. While a larger pitch minimizes crosstalk, it puts a limitation on the on-chip footprint area and the achievable on-chip packing density which is another important figure-of-merit for us. Note that while the FOM graph in Fig. 1(d) shifts towards the right for an increasing width of the waveguide illustrating an increase in the propagation length, the coupling length for a given pitch decreases with the increasing width of the waveguide as shown in Fig. 1(e). This suggests the requirement of higher pitch for wider waveguides for minimizing crosstalk noise, thus increasing the footprint area and reducing the on-chip packing density.

**Realizing Majority Logic Using Surface-Plasmon-Polariton (SPP) Wave**
The functionality of a majority logic gate with *N* (odd) number of inputs is to return a true output if and only if more than half of its inputs (*N/2*) are true. For a SPP wave propagating through a MIM waveguide, the transverse electric field component $E_y$ can be approximated as an exponentially decaying harmonic wave $E_i = E_0 e^{-k_y y} e^{i(k_x x - \omega t + \phi_i)}$ where $E_0$ depends on the strength of the optical or electrical stimulus at the point of excitation and the frequency of the wave $\omega$ is related to the excitation frequency/wavelength. The phase of the wave $\phi$ may be specified by the source or by an additional phase-shifter introduced to achieve a specific phase-shift. We propose the utilization of the phase of the wave instead of the intensity as the state or computational variable. The information bit is encoded in the phase of the wave. Hence, a phase "$\phi$" represents a logic "1" and a phase "$\phi + \pi$" represents a logic "0". The principle phenomena guiding wave-based computing is interference of waves. In general, consider a plasmonic logic gate with odd number of inputs, allowing the input waves to interfere in a "combiner" region and extracting the resultant wave the output end. In general, if $m$ is the number of inputs with phase $\phi$ and $n$ is the number of inputs with phase $\phi + \pi$, the resultant output SPP wave can be approximated as $E_{out} = (m-n)E_0 e^{-k_y y} e^{i(k_x x - \omega t + \phi)}$ with the peak amplitude depending on the number of inputs with phases $\phi$ and $\phi + \pi$ as $E_{out}^{peak} = (m-n)E_0$. A critical requirement of such wave-computing utilizing the phase as the state variable is to have odd number of inputs. An even number of inputs would result in a complete destructive interference (with no output signal) giving rise to an in-determinant state. For the case of 3-input majority logic, the three input SPP waves with the same frequency and amplitude and a certain phase shift relative to each other can be approximated by $E_i = E_0 e^{-k_y y} e^{i(k_x x - \omega t + \phi_i)}$s, where $i = A, B, C$ relates to the three input waves. The output SPP wave resulting from the interference is then, $E_{out} = E_A + E_B + E_C = E_0 e^{-k_y y} e^{i(k_x x - \omega t)} [e^{i\phi_A} + e^{i\phi_B} + e^{i\phi_C}]$. One of the following four possible scenarios arises:

(i) $\phi_A = \phi_B = \phi_C = \phi$
For the case when all the inputs have the same phase $\phi$ (bit "1"), the resultant SPP wave $E_{out} = 3E_0 e^{-k_y y} e^{i(k_x x - \omega t + \phi)}$ has thrice the amplitude and phase $\phi$ resulting in a Boolean logic "1".

(ii) $\phi_A = \phi_B = \phi, \phi_C = \phi + \pi$
When two of the inputs are in the same phase $\phi$ (bit "1") and one is out of phase $\phi + \pi$ (bit "0"), the output SPP wave $E_{out} = E_0 e^{-k_y y} e^{i(k_x x - \omega t + \phi)}$ has the same amplitude as the inputs and phase $\phi$ resulting in a Boolean logic "1".

(iii) $\phi_A = \phi_B = \phi + \pi, \phi_C = \phi$
When two of the inputs are in the same phase $\phi + \pi$ (bit "0") and one is out of phase $\phi$ (bit "1"), the output SPP wave $E_{out} = -E_0 e^{-k_y y} e^{i(k_x x - \omega t + \phi)}$ has the same amplitude as the inputs and phase $\phi + \pi$ resulting in a Boolean logic "0".

(i) $\phi_A = \phi_B = \phi_C = \phi + \pi$
For the case when all the inputs have the same phase $\phi + \pi$ (bit "0"), the resultant SPP wave $E_{out} = -3E_0 e^{-k_y y} e^{i(k_x x - \omega t + \phi)}$ has thrice the amplitude and phase $\phi + \pi$ resulting in a Boolean logic "0".

**Single Stage 3-input Plasmonic Majority Logic Gate**

Next, we investigate a single stage 3-input plasmonic majority logic gate structure illustrated in Figs. 2(a, b). Coherent SPP waves are injected in three parallel input metal-slot waveguides using a standard mode source. The standard mode source in our simulations operate at a wavelength of $\lambda_0 = 1.55 \, \mu m$ (corresponding to a frequency of 193 THz) and inject a fundamental guided mode into the three plasmonic waveguides. The phase of the mode source is specified to be either 0° (from here on referred to as $\phi$) or 180° ($\phi + \pi$) for simulating a logic 1 or 0, respectively. Note that here we rely on the phase of the SPP wave as the computational or state variable. We choose the input waveguide width ($w_{in}$) to be 60 nm and height ($h$) 100 nm. The effective refractive index of the injected SPP mode is calculated to be 1.79+0.0231i which results in a considerable range of usable propagation length of $L_P$ = 5.31 µm (around 6 times the wavelength of the propagating SPP wave $\lambda_{SPP}$ and 5 times the length of the majority gate ~ 1 µm) and a sufficiently high degree of confinement of $\delta$ = 22 nm. The length of input and output waveguide regions ($x_1$) are 200 nm long. The center-to-center distance (pitch $p$) between the adjacent input waveguides in the input region is chosen as 360 nm which corresponds to a coupling length $L_c$~ 10 µm. Since the coupling length is $L_c$ is almost an order of magnitude higher than the length of the majority gate (~ 1 µm), this choice of pitch size provides a good trade-off between crosstalk and on-chip packing density. The combiner region consisting of two bends where the SPPs merge into an output waveguide as shown schematically in Fig. 2(b). To ensure smooth merging of the waveguides with minimum dissipative and backpropagation loss, we choose the merging angle to be 35° between the waveguides which corresponds to a 500 nm long combiner region ($x_2$). The cross-sectional dimensions of the waveguide (60 nm x 100 nm), pitch (360 nm) and merging angle (35°) chosen in this work are comparable to that used by Pan, D. *et. al.* [20] for MIM waveguides resulting in a similar propagation and coupling length. However, while the authors of ref. [20] relied on the intensity of the SPP wave for designing CMOS-oriented logic gates, here we utilize the phase of the SPP wave as the state variable to design majority logic gate.

As the three input waveguides inject power into a single output waveguide, a considerable backflow can occur due to reflection from the merging point. To improve transmission, we increase the gap width of the output waveguide. Numerical simulations (see supplementary section S3 for results) show that the transmitted power through the 3-input junction increases with the increase in the gap width of the output waveguide. However, to avoid mode splitting and large increase in the required pitch for the next stage, we choose a moderately increased output gap width of 120 nm. To better elucidate the result, we resort to an approach of impedance matching put forward by Cai *et. al*[46] which shows an increase in the transmission due to an increase in the impedance of the waveguide (see supplementary section S3 for further details and results). The SPPs propagating through the two side arms of the majority gate have to cover an extra distance introduced by the bends. Since we rely purely on a constructive (phase difference = 0) or destructive interference (phase difference = $\pi$) between the input SPP waves, we compensate the path difference $\delta l$ between the middle and side arms and make all the three inputs equal in phase and strength by shifting the position of the mode source for the middle input waveguide to the left by a distance $\delta l = \left( \frac{p}{\sin\left[tan^{-1}\left(\frac{p}{x_2}\right)\right]} - x_2 \right)$ as shown in Fig. 2(b).

Fig. 2(c) shows the numerical simulation result for a 3-input plasmonic majority gate for all $2^3$ possible input combinations in terms of the time-domain electric field component $E_Y$ at the output, integrated over the cross-section of the output waveguide and normalized to the total source electric field and cross-sectional area of the output waveguide (see method section for details). The input logic combinations (1,1,1), (1,1,0), (1,0,1) and (0,1,1) which correspond to majority of the input being logic 1, i.e., SPP waves having a phase $\phi$, result in logic 1 as the output, i.e., an output SPP wave having a phase $\phi$. Similarly, the input combinations (0,0,0), (0,0,1), (0,1,0) and (1,0,0) result in an output SPP wave with phase $\phi + \pi$ depicting a logic 0 as the output. In addition to generating Boolean outputs, the proposed majority logic gate also has the capability to distinguish between a *strong* and a *weak* majority. As highlighted by the simulation results in Fig. 2(c), an input logic combination of (1,1,1) corresponding to all the SPP waves having a phase $\phi$, gives rise to maximum constructive interference resulting in a high amplitude output with the phase $\phi$, i.e., a strong 1. On the other hand, for an input logic combination of (1,1,0), two of the SPP waves having phase $\phi$ and $\phi + \pi$ undergo destructive interference resulting in a low amplitude output with the phase $\phi$, i.e., a weak logic 1. Fig. 2(d) shows the peak values of the integrated electric field component $E_Y$ at the output, normalized to the source electric field, for different combinations of the input phases. As can be seen, the output of a single stage 3-input majority gate has four different levels denoting a combination of Boolean output 1 and 0 and the strength of the majority. Also, the normalized peak output $E_Y$ for strong majority (all inputs in phase) is 3 times stronger than that of the weak majority (two in phase, one out of phase) as explained earlier. To further study the propagation and interference of SPP waves in the majority logic structure, we plot the time-lapse simulation results in terms of the distribution of the electric field component $E_Y$ in the x-y plane at different snapshots in time for 4 different input combinations (1,1,1), (1,0,1), (1,0,0) and (0,0,0) as seen in Fig. 2(e) (see supplementary section S4 for further simulation results).

Note that the majority gate also has the capability to perform "AND" and "OR" operation if one of the inputs is used as a control input. Additionally, due to the wave nature of this computation scheme, an inverter (INV) operation can be simply implemented using a waveguide of length equal to half of the SPP wavelength. This set of logic primitives allows to effectively map most practical arithmetic functions, even when they cannot be directly matched well to just a cascaded majority logic structure. The multi-level output of the majority gate depicts a combination of Boolean output 1 and 0 and the strength of the majority. In order to utilize such a majority gate for Boolean computation, one needs to renormalize the output before feeding it to the next stage. However, here we strive to utilize this multi-level output to our advantage for non-Boolean computing. Since each of the stages in a cascaded structure (see Figs. 3(a)) performs the dual functionality of Boolean output and strength of majority, the overall final result will display the Boolean logic output 1 or 0 and the overall strength the majority of all the inputs.

## 2-Stage Cascaded Plasmonic Majority Logic Gate

Next, we analyze a 2-stage cascaded majority gate structure shown in Fig. 3(a). The first stage consists of three '3-input majority gates', each of whose outputs are combined in a majority gate fashion to form the second stage (see supplementary section S5 for the dimensional scaling implemented at each stage). In our numerical simulations, all nine inputs are excited with the same stimulus in terms of source amplitude. The phase of the injected SPP waves are chosen to be either $\phi$ or $\phi + \pi$ representing logic 1 and 0 respectively. The final output is monitored at the end of the second stage as indicated in Fig. 3(a). As mentioned earlier, the propagating SPP waves can be considered as exponentially decaying harmonic waves with the same frequency and amplitude and a certain phase shift relative to each other and can be approximated as $E_i = E_0 e^{-k_y y} e^{i(k_x x - \omega t + \phi_i)}$s, where $i = 1, \ldots, 9$ relates to the nine input waves for each of the distinct arms. The resultant output SPP wave due to wave interference can be written as $E_{out} = E_1 + E_2 + \cdots + E_8 + E_9 = E_0 e^{-k_y y} e^{i(k_x x - \omega t)} [e^{i\phi_1} + e^{i\phi_2} + \cdots + e^{i\phi_8} + e^{i\phi_9}]$. There are $2^9$ possible input combinations for such a 2-stage cascaded majority logic gate, however here we only concentrate on 10 representative cases as highlighted in Figs. 3(b) and (c) that capture all the possible combinations of input phases required for studying the cascaded majority gate structure.

Fig. 3(b) shows the numerical simulation results for the 10 representative input combinations in terms of the time-domain normalized integrated electric field component E_Y at the output. As explained earlier, the resultant output SPP wave can be approximated as $E_{out} = (m - n) E_0 e^{-k_y y} e^{i(k_x x - \omega t + \phi)}$, $m$ being the number of inputs with phase $\phi$ and $n$ is the number of inputs with phase $\phi + \pi$. For the case when all the inputs are logic "1", i.e., having the same phase $\phi$, the resultant SPP wave $E_{out} = 9 E_0 e^{-k_y y} e^{i(k_x x - \omega t + \phi)}$ has nine time the amplitude and phase $\phi$ resulting in a Boolean logic "1". In the extreme opposite case when all the inputs are logic "0", i.e., having the same phase $\phi + \pi$, the resultant SPP wave $E_{out} = -9 E_0 e^{-k_y y} e^{i(k_x x - \omega t + \phi)}$ has nine time the amplitude but phase $\phi + \pi$ resulting in a Boolean logic "0". The peak amplitude of the output electric field depends on the number of inputs with phases $\phi$ (m) and $\phi + \pi$ (n) and varies as $E_{out}^{peak} = (m - n) E_0$ as seen in Fig. 3(b, c). Overall, we obtain 10 different levels of output electric field component E_Y which directly corresponds to the strength of the majority of the input. The strongest majorities where all the 9 inputs have the same phase $\phi$ or $\phi + \pi$ produce the highest magnitude of output amplitude with opposite sign as indicated in Fig. 3(c). Likewise, the weakest majorities where only 5 inputs have phase $\phi$ while the remaining 4 have phase $\phi + \pi$ or vice-versa produce the lowest magnitude of output amplitude. The time-lapse simulation results in terms of the distribution of the electric field component E_Y in the x-y plane for 4 representative input combinations (111,111,111), (111,110,110), (000,001,001) and (000,000,000) are shown in Fig. 3(d) depicting the propagation and interference of the SPP waves (see supplementary section S6 for further simulation results).

## Referencing Technique for Detection

As mentioned earlier, the binary data is encoded in the phase of the excited SPP wave (phase $\phi$ representing logic 1 and phase ($\phi + \pi$) representing logic 0) which we use as the state variable for computing. Hence, after wave interference, the relevant parameter to extract is the phase of the output SPP wave which gives the Boolean output of 1 or 0. However, it may be challenging to devise a precise tera-Hertz phase-detection scheme for our plasmonic logic gate operating at 193 THz (see dispersion plot in Fig. 1(b)). In addition, our proposed plasmonic majority gate also displays a non-Boolean characteristic in terms of indicating the strength of the majority which further adds to the expressive power of the plasmonic majority gate. As such, in this work, we propose a novel approach to extract both the amplitude and phase information from the output by using a "referencing" technique. An illustration for using the referencing technique at the end of the 2-stage cascaded structure is shown in Fig. 4(a). In addition to the 9 input SPP waves, we also inject a reference signal $E_{ref}$ that merges at the output of the second stage. We choose the gap width of the reference waveguide to be *3w*. We adjust the amplitude and the phase of the injected reference signal $E_{ref} \approx 9E_0 e^{-k_y y} e^{i(k_x x - \omega t + \phi)}$ to match the output of the cascaded gate for the case of the strongest majority logic 1 (the case of all inputs 1 in Fig. 3(b)). This can be done by using higher excitation power for the reference source and accounting for the total path delay for the reference signal. Fig. 4(b) shows the numerical simulation result for the 2-stage cascaded majority with reference for the 10 representative input combinations in terms of the time-domain normalized integrated electric field component E$_Y$ at the output. Since the reference signal has been adjusted to have a phase $\phi$, all the input combinations having majority of the input as logic 1, i.e., SPP waves having a phase $\phi$, result in a constructive interference while all majority 0 cases having phase $\phi + \pi$ result in destructive interference. However, since we set the peak amplitude of the reference signal to be $E_{ref}^{peak} \geq 9E_0$, we get a Boolean output of logic 1 for all the 10 cases (output SPP waves having phase $\phi$), but with varying levels of output amplitude of the electric field component E$_Y$ as indicated in figs. 4(b) and (c). In contrast to Figs. 3(c) and (d), in this case the strongest majority with 9 inputs having phase $\phi$ produces the highest magnitude of output electric field E$_Y$ and maximum transmission of output power while the strongest majority with 9 inputs having phase $\phi + \pi$ produces the lowest magnitude of output E$_Y$ and minimum power transmission as indicated in Fig. 4(c) and (d). Thus, the referencing technique directly translates the information encoded in the amplitude and phase of the output SPP wave into the intensity of the output electric field or the output power as illustrated in fig. 4(c) and (d). We can further define the peak amplitude of the output electric field or the output power transmitted for the case when only the reference signal is present as the threshold level. As such, anything above the defined threshold can be considered as a logic 1 while anything below gives a logic 0. Fig. 4(e) shows the time-lapse simulation results in terms of the distribution of the electric field component E$_Y$ in the x-y plane for 4 representative input combinations (111,111,111), (111,110,110), (000,001,001) and (000,000,000), depicting the propagation and interference of the SPP waves along with the reference signal (see supplementary section S7 for further simulation results).

## Discussion
**Number of Cascadable Stages**

While it is highly desired to have a multi-staged cascaded plasmonic logic without intermediary signal conversion between plasmon and charge domain, the propagation loss of SPP puts a limitation on the number of feasible cascaded stages. As shown in Fig. 5(a) (and Fig. 4(b) in supplementary information), the size of majority logic gate increases with the number of stages from an estimated value of 0.636 $\mu m^2$ for the first stage to 4.66 $\mu m^2$ and 38.24 $\mu m^2$ for the second and third stage, respectively. The increase in the path-length travelled by the SPP compared to the propagation length $L_P$ increases the transmission loss from around 30% in the first stage to more than 50% in the third stage. Hence, it is inefficient to go beyond the third stage without using either amplifiers to boost the signal amplitude or convert plasmonic signal to voltage signal at the end of the third stage. An additional constraint comes from distinction or separation between output levels after referencing. The number of output levels (amplitude of electric field $E_Y$ or transmitted power) increases with the number of stages and input, from 4 in $1^{st}$ stage to 10 in the $2^{nd}$ and so on. Fig. 5(b) shows the range of amplitude of the output electric field $E_Y$ for logic 1 and 0 obtained at the end of each stage. Note that the range of output for both logic 1 and 0 decreases due the propagation loss from one stage to the next. Hence, even though the referencing technique will translate the information encoded in the amplitude and phase of the output SPP wave into electric field intensity, it will be difficult to separate or distinguish between the output levels for logic 1 and 0 as they get closer to the threshold level (case of weakest majority). We further investigate the possibility of separation of states above and below the threshold level (corresponding to logic 1 and 0) by plotting the resolution as a function of the number of stages shown in Fig. 5(c). We define the resolution as the difference between the minimum value of peak output $E_Y$ for logic 1 and the maximum value of peak output $E_Y$ for logic 0 (case of weakest majority outputs), $resolution = \Delta E_{Y,out} = E_{Y,min}^{logic\ 1} - E_{Y,max}^{logic\ 0}$.

**Comparison with other proposals for plasmonic logic and device**

Our work on majority logic gate design focuses on metal-insulator-metal (MIM) wave-guiding configuration for high field confinement capability (in nanometers) that aids the designing of nanoscale logic devices. Recently, a work on plasmonic majority gate using long-range dielectric-loaded surface plasmon polariton (LR-DLSPP) waveguide has also been proposed [47]. While the latter supports long range propagation of plasmon polaritons ($L_P$ ~ 3 mm), it has a rather poor confinement (lateral mode width, $\delta =$ ~ 1.6µm) when compared to the high degree of confinement ($\delta$ = 22 nm) achieved in our proposal. With a waveguide length and width of 14 µm and 5.5 µm, respectively, in addition to a pitch of 0.5 µm, the single-stage 3-input majority gate proposed in [47] requires an area of 77 µm². On the contrary, our single-stage MIM-based 3-input majority gate has a highly improved overall area of only 0.636 µm². The cross-sectional dimensions of the waveguide (60 nm x 100 nm) chosen in this work are comparable to that used by Pan et. al. [20] resulting in a similar coupling length of around 12 µm for a chosen pitch of 300 nm. Similar investigations using MIM geometry include a 200 nm x 100 nm air groove based cascaded XOR gate resulting in an all-optical logic parity checker where an overall minimum feature size of 15 µm for the logic device was achieved [48]. A 320 nm x 300 nm dielectric crossed waveguide structure enabling all-optical NOT, AND, OR, and XOR gate with lateral area of 200 µm² [49] and a half-adder structure of area 280 µm² was proposed. The all-optical realization of XNOR, XOR, NOT, and OR logic gates by Fu. et. al. used 100 x 100 nm slot waveguide [50]. However, the latter used a pitch of 2 µm resulting in the lateral dimension of the logic gate to be close to 5 µm. The experimental demonstration of OR/NOT/NOR using cascaded plasmonic logic using Ag nanowires (NWs) [18,19] reports using 200 - 350 nm diameters NWs with a few to tens of micron length. Recently, the surface plasmon two-mode interference (SPTMI) coupler featuring a silicon core (width 0.22 – 0.48 µm), a silver upper and lower cladding, and GaAsInP left and right cladding has been proposed as an alternative for low power consumption that realizes the complete set of logic operations. However, the device dimensions still remain comparatively large: 1x2 SPTMI requiring coupling length of 93.1 µm results in a total length of 143.06 um [51], 2x2 SPTMI single stage has a coupling length 38.4 µm and total length of 88.34 µm [52], and a two-stage cascaded structure has a coupling length of 92.35 µm resulting in total length of device equal to 284.58 µm [53].

**Application Illustration**
Due to high throughput, the proposed cascaded plasmonic majority gate can be of great use in highly parallel real-time signal and data processing applications. One good illustration of this usage is present in the non-boolean decision making process of a pattern recognition system. The non-boolean decision making process involves counting the number of matches and mismatches and determining the degree of match or mismatch between the input and the reference pattern. We believe the proposed cascadable plasmonic majority logic gate can find a direct utility here. The patterns can be considered as binary valued matrices with black and white pixels represented as logic "1" and "0", respectively. Using the majority voting capability of the gate along with the referencing technique, it would be possible to count the number of match or mismatch at each stage with the final output portraying an overall match or mismatch between the input and the reference pattern and the degree of match or mismatch found.

In conclusion, using numerical FDTD simulations, we have demonstrated the possibility of building a nanoscale cascadable plasmonic majority gate. We utilize the phase of the SPP wave, instead of the intensity, as the state or computational variable that allows us to exploit the majority voting capability of the device inaccessible to amplitude-based wave computing. We choose the MIM geometry that allows nanometer scale plasmonic mode confinement, sub-micrometer pitch and propagation lengths over several micrometers. The dimensions and pitch of the waveguide chosen are comparable to that used by Pan, D. *et. al.* [20] for MIM waveguides. However, while the authors of ref. [20] relied on the intensity of the SPP wave for designing CMOS-oriented logic gates, here we utilize the phase of the SPP wave as the state variable to design majority logic gate. Comparison with other previous works of plasmonic logic [47-53] reveals that our single-stage MIM-based 3-input majority gate has a highly improved overall area of only 0.636 µm$^2$. Performing 3-D FDTD simulations, we illustrate the majority logic functionality of the proposed gate and its cascadability up to 3 stages. In addition to boolean logic, the proposed plasmonics majority gate is also capable of performing non-boolean computing due to multiple output levels denoting the strength of the majority. We also propose a novel referencing scheme at the output that can translate the information encoded in the phase of the output SPP wave into the intensity of the output electric field. The extremely high throughput of plasmonic based logic finds direct usage in high throughput low latency signal processing applications which are arithmetic-heavy with strict timing requirements, like a pattern recognition system.

## Methods
**Numerical simulation**

We perform full-wave 3-D simulations using the commercially available software Lumerical Solution[24]. The finite difference time domain (FDTD) method is used to solve the fully vectorial Maxwell equation on a discrete spatial (Yee cell) and temporal grid. We use the conformal mesh algorithm with a specified mesh size of 5 nm for regular straight waveguides and a reduced mesh of 2 nm or 2.5 nm for bends. Absorbing boundary conditions based on perfectly matched layers (PML) are used to minimize reflections. We use the mode source to inject the fundamental guided mode into the plasmonic waveguide. The standard mode source injects a broadband Gaussian pulse signal, but the results and the conclusion of the paper remain same when using a narrower-band mode source (see supplementary section S2 and S8 for further details). The phase of the source is specified as either 0 or $180^{o}$ for simulating either a logic 1 or 0. The refractive index of Ag in our simulations is obtained from Palik's Handbook of Optical Constants [41]. We use the electromagnetic field and power monitor (2D x-normal) in Lumerical to record the normalized transmission coefficients at the output. The time-domain electric field results are obtained using the time-domain monitor in Lumerical (2D x-normal for output $E_Y$ and 2D z-normal for time-lapse results). The output electric field component $E_Y$ is calculated as a spatial average of the y-component of the electric field distribution, obtained from the time-domain monitor, over the cross-section of the output waveguide.

## References

1       Zhirnov, V. V., Cavin, R. K., Hutchby, J. A. & Bourianoff, G. I. Limits to binary logic switch scaling-a gedanken model. *Proc. Ieee* **91**, 1934-1939 (2003).
2       Nikonov, D. & Young, I. Overview of beyond-cmos devices and a uniform methodology for their benchmarking. *Proc. Ieee* **101**, 2498-2533 (2013).
3       Caulfield, H. J. & Dolev, S. Why future supercomputing requires optics. *Nature Photonics* **4**, 261-263 (2010).
4       Barnes, W. L., Dereux, A. & Ebbesen, T. W. Surface plasmon subwavelength optics. *Nature* **424**, 824-830 (2003).
5       Ozbay, E. Plasmonics: merging photonics and electronics at nanoscale dimensions. *science* **311**, 189-193 (2006).
6       Ditlbacher, H. *et al.* Silver nanowires as surface plasmon resonators. *Physical review letters* **95**, 257403 (2005).
7       Sanders, A. W. *et al.* Observation of plasmon propagation, redirection, and fan-out in silver nanowires. *Nano letters* **6**, 1822-1826 (2006).
8       Yan, R., Pausauskie, P., Huang, J. & Yang, P. Direct photonic–plasmonic coupling and routing in single nanowires. *Proceedings of the National Academy of Sciences* **106**, 21045-21050 (2009).

9   Chen, X.-W., Sandoghdar, V. & Agio, M. Highly efficient interfacing of guided plasmons and photons in nanowires. *Nano letters* **9**, 3756-3761 (2009).
10  Fang, Y., Wei, H., Hao, F., Nordlander, P. & Xu, H. Remote-excitation surface-enhanced Raman scattering using propagating Ag nanowire plasmons. *Nano letters* **9**, 2049-2053 (2009).
11  Holmgaard, T. & Bozhevolnyi, S. I. Theoretical analysis of dielectric-loaded surface plasmon-polariton waveguides. *Physical Review B* **75**, 245405 (2007).
12  Veronis, G. & Fan, S. Guided subwavelength plasmonic mode supported by a slot in a thin metal film. *Optics letters* **30**, 3359-3361 (2005).
13  Krasavin, A. & Zayats, A. Passive photonic elements based on dielectric-loaded surface plasmon polariton waveguides. *Applied Physics Letters* **90**, 211101 (2007).
14  Bozhevolnyi, S. I., Volkov, V. S., Devaux, E., Laluet, J.-Y. & Ebbesen, T. W. Channel plasmon subwavelength waveguide components including interferometers and ring resonators. *Nature* **440**, 508-511 (2006).
15  Dionne, J. A., Diest, K., Sweatlock, L. A. & Atwater, H. A. PlasMOStor: a metal– oxide– Si field effect plasmonic modulator. *Nano Letters* **9**, 897-902 (2009).
16  Cai, W., White, J. S. & Brongersma, M. L. Compact, high-speed and power-efficient electrooptic plasmonic modulators. *Nano letters* **9**, 4403-4411 (2009).
17  MacDonald, K. F., Sámson, Z. L., Stockman, M. I. & Zheludev, N. I. Ultrafast active plasmonics. *Nature Photonics* **3**, 55-58 (2009).
18  Wei, H., Wang, Z., Tian, X., Käll, M. & Xu, H. Cascaded logic gates in nanophotonic plasmon networks. *Nature communications* **2**, 387 (2011).
19  Wei, H. *et al.* Quantum dot-based local field imaging reveals plasmon-based interferometric logic in silver nanowire networks. *Nano letters* **11**, 471-475 (2010).
20  Pan, D., Wei, H. & Xu, H. Optical interferometric logic gates based on metal slot waveguide network realizing whole fundamental logic operations. *Opt Express* **21**, 9556-9562, doi:10.1364/OE.21.009556 (2013).
21  Amarú, L., Gaillardon, P.-E. & De Micheli, G. in *Proceedings of the 51st Annual Design Automation Conference.*  1-6 (ACM).
22  Dutta, S. *et al.* Non-volatile clocked spin wave interconnect for beyond-CMOS nanomagnet pipelines. *Scientific reports* **5** (2015).
23  Zografos, O. *et al.* Non-volatile spin wave majority gate at the nanoscale. *AIP Advances* **7**, 056020, doi:10.1063/1.4975693 (2017).
24  Lumerical, F. Solutions. *Web source [*https://www/*. lumerical. com/tcad-products/fdtd/]* (2012).
25  Ebbesen, T. W., Genet, C. & Bozhevolnyi, S. I. Surface-plasmon circuitry. *Physics Today* (2008).
26  Maksymov, I. S. & Kivshar, Y. S. Broadband light coupling to dielectric slot waveguides with tapered plasmonic nanoantennas. *Optics letters* **38**, 4853-4856 (2013).
27  Panchenko, E., James, T. D. & Roberts, A. Modified stripe waveguide design for plasmonic input port structures. *Journal of Nanophotonics* **10**, 016019-016019 (2016).
28  Koller, D. *et al.* Organic plasmon-emitting diode. *Nature Photonics* **2**, 684-687 (2008).

29  Neutens, P., Lagae, L., Borghs, G. & Van Dorpe, P. Electrical excitation of confined surface plasmon polaritons in metallic slot waveguides. *Nano letters* **10**, 1429-1432 (2010).
30  Walters, R. J., van Loon, R. V., Brunets, I., Schmitz, J. & Polman, A. A silicon-based electrical source of surface plasmon polaritons. *Nature Materials* **9**, 21-25 (2010).
31  Cazier, N. *et al.* Electrical excitation of waveguided surface plasmons by a light-emitting tunneling optical gap antenna. *Optics express* **24**, 3873-3884 (2016).
32  Uskov, A. V., Khurgin, J. B., Protsenko, I. E., Smetanin, I. V. & Bouhelier, A. Excitation of plasmonic nanoantennas by nonresonant and resonant electron tunnelling. *Nanoscale* **8**, 14573-14579 (2016).
33  Kern, J. *et al.* Electrically driven optical antennas. *Nature Photonics* **9**, 582-586 (2015).
34  Knight, M. W., Sobhani, H., Nordlander, P. & Halas, N. J. Photodetection with active optical antennas. *Science* **332**, 702-704 (2011).
35  Falk, A. L. *et al.* Near-field electrical detection of optical plasmons and single-plasmon sources. *Nature Physics* **5**, 475-479 (2009).
36  Neutens, P., Van Dorpe, P., De Vlaminck, I., Lagae, L. & Borghs, G. Electrical detection of confined gap plasmons in metal–insulator–metal waveguides. *Nature Photonics* **3**, 283-286 (2009).
37  Ly-Gagnon, D.-S. *et al.* Routing and photodetection in subwavelength plasmonic slot waveguides. *Nanophotonics* **1**, 9-16 (2012).
38  Zia, R., Selker, M. D., Catrysse, P. B. & Brongersma, M. L. Geometries and materials for subwavelength surface plasmon modes. *JOSA A* **21**, 2442-2446 (2004).
39  Economou, E. Surface plasmons in thin films. *Physical review* **182**, 539 (1969).
40  Bozhevolnyi, S. I. & Jung, J. Scaling for gap plasmon based waveguides. *Optics express* **16**, 2676-2684 (2008).
41  Palik, E. D. *Handbook of optical constants of solids*. Vol. 3 (Academic press, 1998).
42  Khurgin, J. B. & Sun, G. Scaling of losses with size and wavelength in nanoplasmonics and metamaterials. *Applied Physics Letters* **99**, 211106 (2011).
43  Maier, S. A. *Plasmonics: fundamentals and applications*.  (Springer Science & Business Media, 2007).
44  Oulton, R., Bartal, G., Pile, D. & Zhang, X. Confinement and propagation characteristics of subwavelength plasmonic modes. *New Journal of Physics* **10**, 105018 (2008).
45  Dastmalchi, B., Tassin, P., Koschny, T. & Soukoulis, C. M. A new perspective on plasmonics: confinement and propagation length of surface plasmons for different materials and geometries. *Advanced Optical Materials* **4**, 177-184 (2016).
46  Cai, W., Shin, W., Fan, S. & Brongersma, M. L. Elements for Plasmonic Nanocircuits with Three-Dimensional Slot Waveguides. *Advanced materials* **22**, 5120-5124 (2010).
47  Doevenspeck, J. *et al.* Design and simulation of plasmonic interference-based majority gate. *AIP Advances* **7**, 065116 (2017).
48  Wang, F. *et al.* Nanoscale on-chip all-optical logic parity checker in integrated plasmonic circuits in optical communication range. *Scientific reports* **6**, 24433 (2016).
49  Birr, T., Zywietz, U., Chhantyal, P., Chichkov, B. N. & Reinhardt, C. Ultrafast surface plasmon-polariton logic gates and half-adder. *Optics express* **23**, 31755-31765 (2015).

50  Fu, Y. *et al.* All-optical logic gates based on nanoscale plasmonic slot waveguides. *Nano letters* **12**, 5784-5790 (2012).
51  Sahu, P. P. Theoretical investigation of all optical switch based on compact surface plasmonic two mode interference coupler. *Journal of Lightwave Technology* **34**, 1300-1305 (2016).
52  Gogoi, N. & Sahu, P. P. All-optical compact surface plasmonic two-mode interference device for optical logic gate operation. *Applied optics* **54**, 1051-1057 (2015).
53  Gogoi, N. & Sahu, P. P. All-optical surface plasmonic universal logic gate devices. *Plasmonics* **11**, 1537-1542 (2016).

## Author Contributions

S.D., F.C. and A.N. developed the main idea. O.Z., S.G., I.R. and B.S. participated in useful discussions and further development of the idea. S.D. performed all the simulations with help from S.G. and O.Z. All authors discussed the results, agreed to the conclusions of the paper and contributed to the writing of the manuscript.

**Competing financial interests:** The authors declare that they have no competing interests.

**Figures**

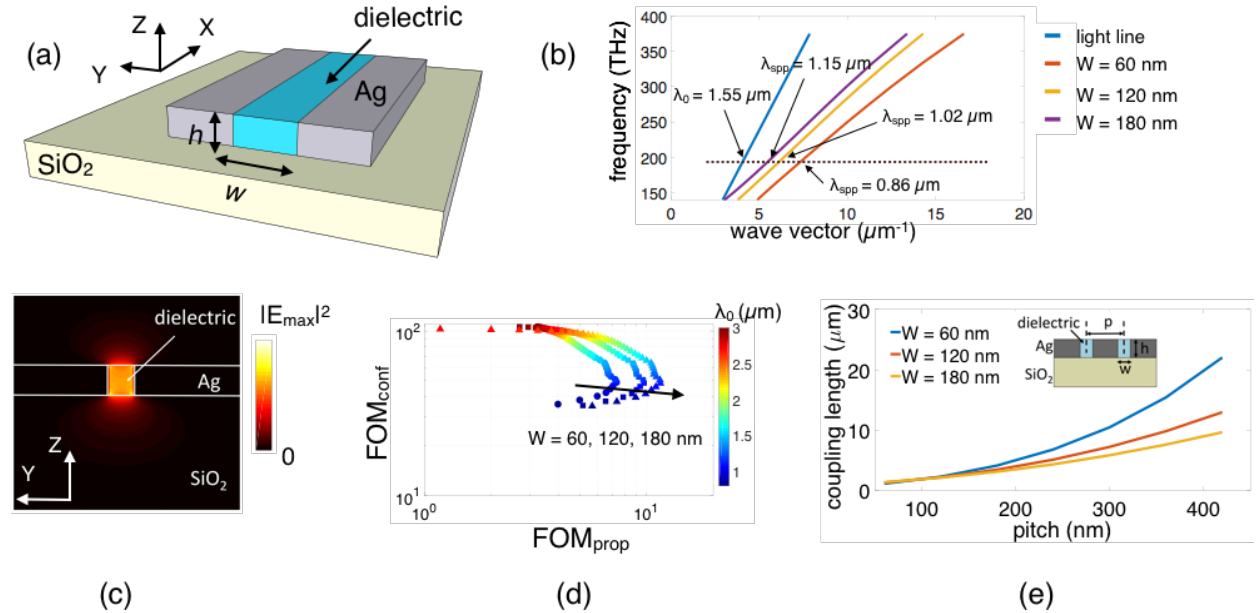

(a) (b)

(c) (d) (e)

Figure 1. (a) Schematic of the metal-slot MIM plasmonic waveguide consisting of Ag on top of SiO$_2$ substrate and a dielectric of 1.5 refractive index, acting as a channel for transmission of information. (b) Calculated dispersion relation of the slot waveguide for gap widths of 60, 120 and 180 nm. (c) Fundamental plasmonic mode in a 60 nm wide metal-slot waveguide at $\lambda_0 = 1.55\ \mu m$. (d) Two dimensional figure of merit (FOM) graph as a function of the wavelength $\lambda_0$ illustrating for our exploration space a good trade-off between the propagation and confinement at the chosen excitation wavelength of $\lambda_0 = 1.55\ \mu m$ (e) Plot showing coupling length $L_p$ as a function of the waveguide pitch $p$ for gap widths of 60, 120 and 180 nm, illustrating the trade-off between crosstalk and on-chip packing density.

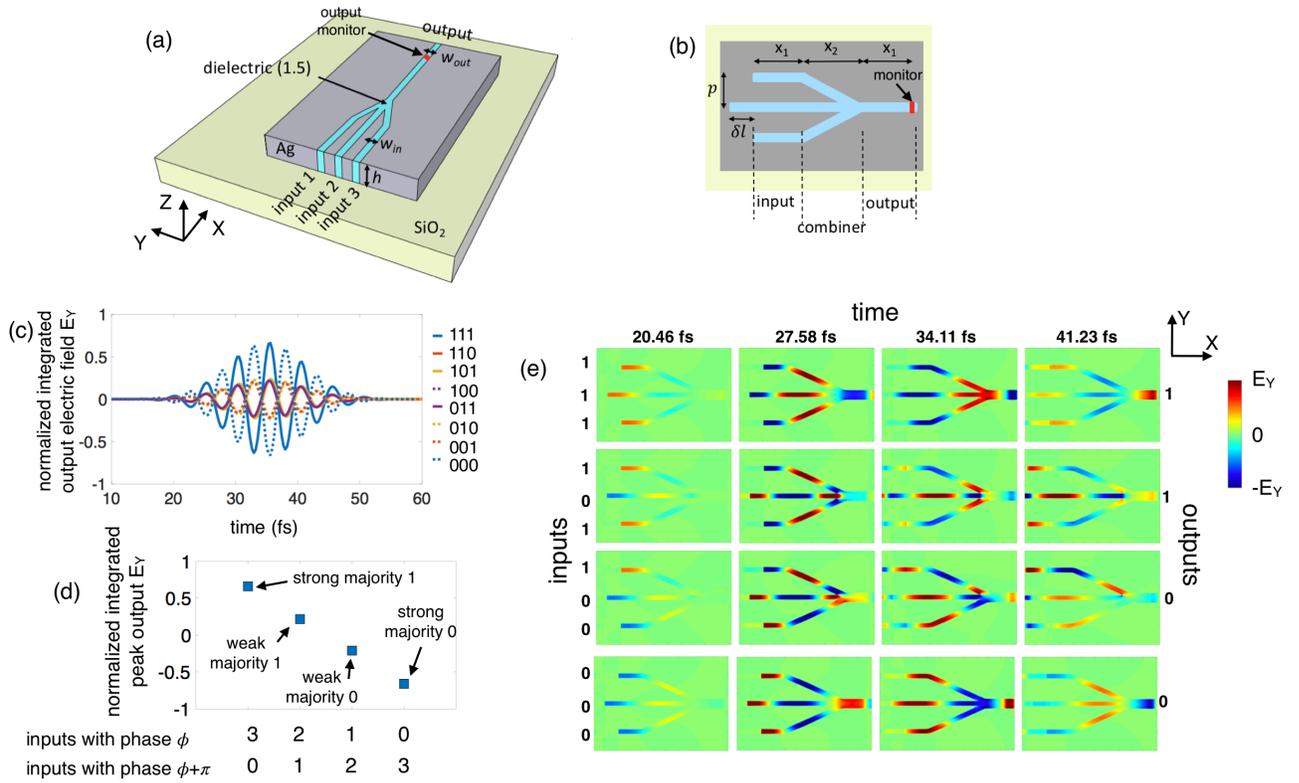

Figure 2. (a, b) Illustration of a single stage 3-input plasmonic majority logic gate. (c) Simulation result for a 3-input plasmonic majority gate for $2^3$ input combinations in terms of the time-domain electric field component $E_Y$ at the output, normalized to the total source electric field and integrated over the cross-section of the output waveguide. (d) Calculated peak values of the normalized integrated electric field component $E_Y$ at the output for different combinations of the input phases. (e) Time-lapse simulation results in terms of the distribution of $E_Y$ in the x-y plane showing the propagation and interference of the SPP waves.

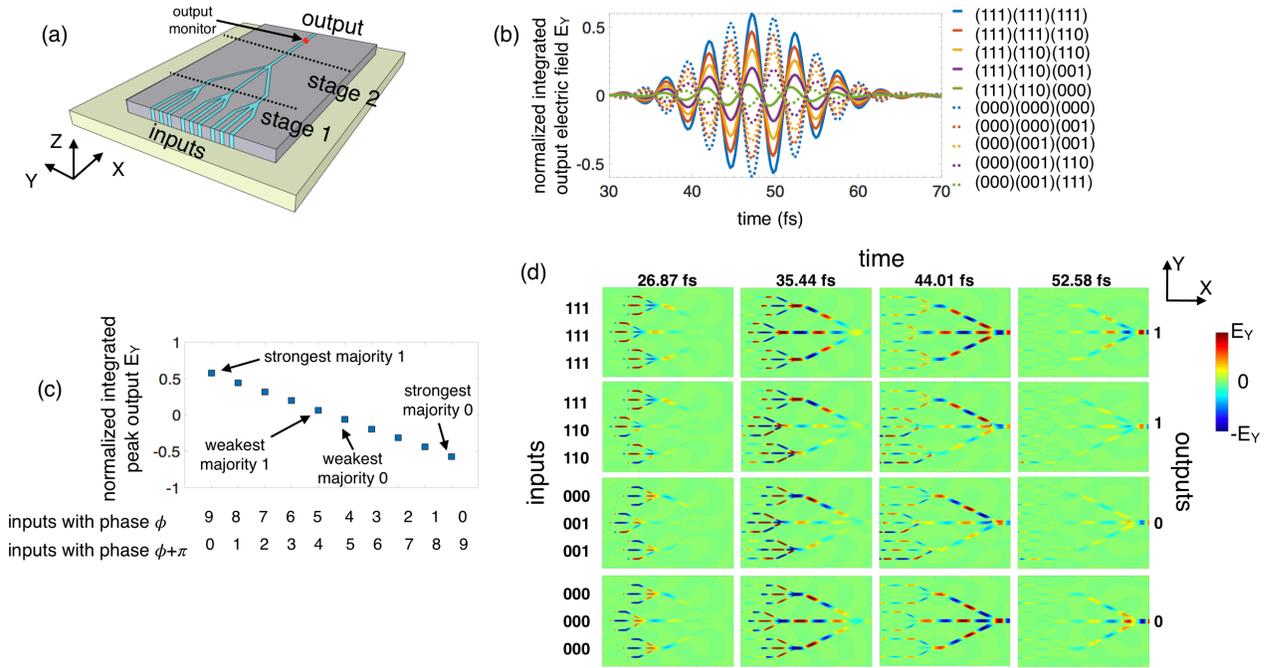

Figure 3. (a) Illustration of a 2-stage cascaded plasmonic majority logic gate. (b) Simulation results for the 10 representative input phase combinations in terms of the time-domain electric field component $E_Y$ at the output, normalized to the total source electric field and integrated over the cross-section of the output waveguide. (c) Calculated peak values of the normalized integrated electric field component $E_Y$ at the output for different combinations of the input phases. (d) Time-lapse simulation results in terms of the distribution of $E_Y$ in the x-y plane showing the propagation and interference of the SPP waves.

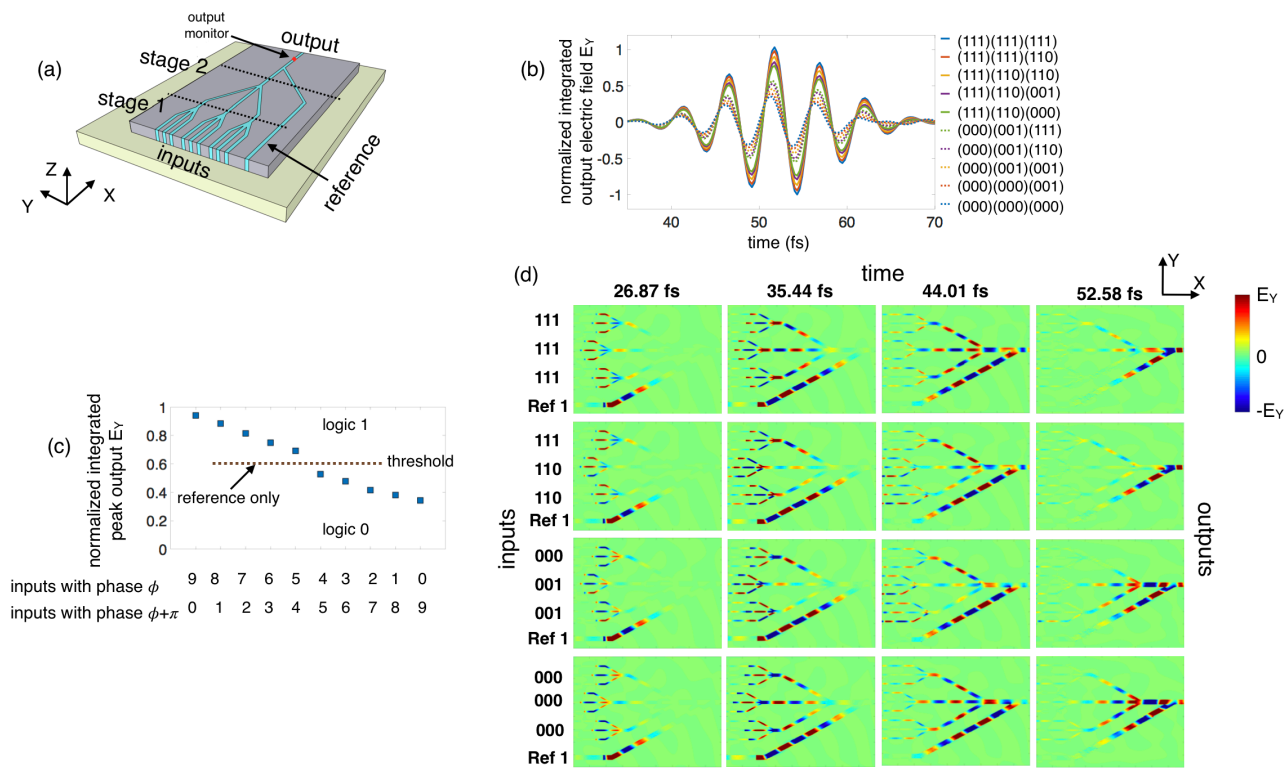

Figure 4. (a) Illustration of a 2-stage cascaded plasmonic majority logic gate with the reference signal. (b) Simulation results for the 10 representative input phase combinations in terms of the time-domain electric field component $E_Y$ at the output, normalized to the total source electric field and integrated over the cross-section of the output waveguide. (c) Calculated peak values of the normalized integrated electric field component $E_Y$ at the output for different combinations of the input phases. The peak amplitude of the output electric field in (c) for the case when only the reference signal is present is defined as the threshold level. (d) Time-lapse simulation results in terms of the distribution of $E_Y$ in the x-y plane showing the propagation and interference of the SPP waves.

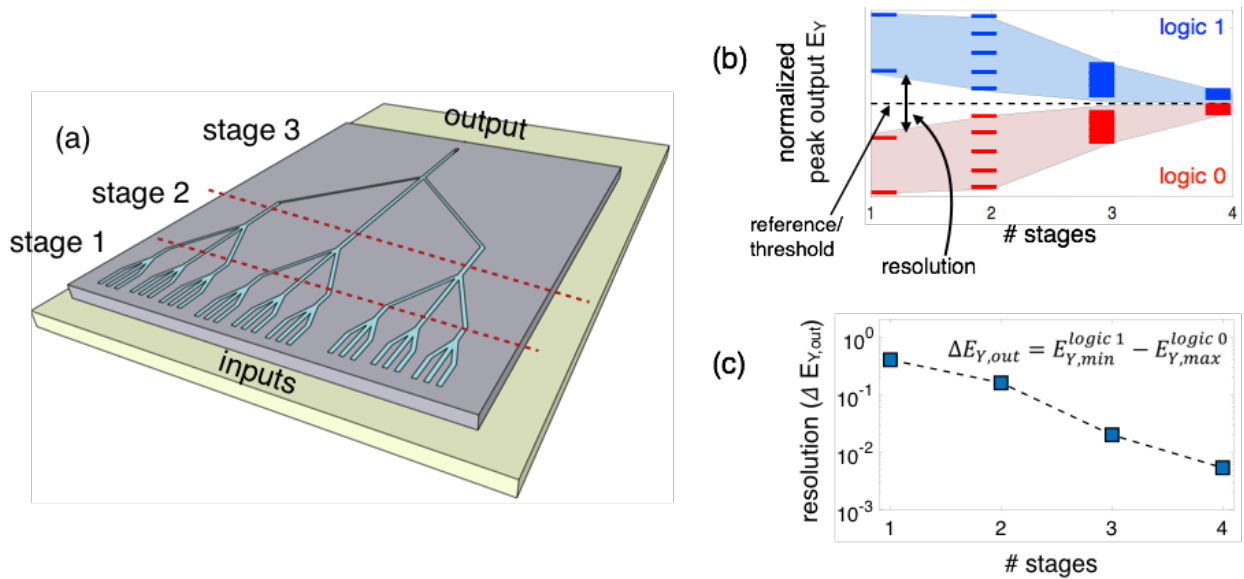

Figure 5. (a) Illustration of a multi-stage cascaded plasmonic majority logic. (b) Plot showing the range of amplitude of the output electric field $E_Y$ for logic 1 and 0 obtained at the end of each stage. The range of output decreases due to propagation loss at each stage. (c) Resolution, defined as the difference between the minimum value of peak output $E_Y$ for logic 1 and the maximum value of peak output $E_Y$ for logic 0 (case of weakest majority outputs), as a function of the number of stages.

# Proposal for nanoscale cascaded plasmonic majority gates for non-boolean computation – Supplementary Information

Sourav Dutta[1*], Odysseas Zografos[2], Surya Gurunarayanan[2], Iuliana Radu[2], Bart Soree[2], Francky Catthoor[2], Azad Naeemi[1]

[1]School of Electrical and Computer Engineering, Georgia Institute of Technology, Atlanta, Georgia 30332, USA
[2]IMEC, Leuven, Belgium

*Correspondence to sdutta38@gatech.edu

## S1. Excitation of SPP

On-chip propagating SPPs can be launched in several ways - optically by focusing external laser radiation on grating couplers [1-3] or electrically using light-emitting diodes [4,5], Si-based electrical source [6] and electron tunneling [7-9]. Phase coherency is an important requirement for our 3-input majority logic operation. Currently, phase coherent waves are not yet demonstrated for two independent emitters. However, provided sufficient input power, by splitting the output of a single emitter into three components and injecting them into the input waveguides, phase coherency can be assured. Since we rely on the phase of the SPP wave as the state or computational variable, the information can be written into the phase of wave in each of the three individual waveguides of the majority logic by incorporating phase modulators such as Mach-Zehnder interferometers (MZI) [10] or shift-keying based resonators [11]. However, since the emphasis of this work is on building a nanoscale cascadable plasmonic majority logic, we do not refer to a particular technique of excitation and use the generalized mode source in Lumerical Solutions[12] to inject a fundamental guided mode into the plasmonic waveguide with the phase of the source specified as either 0 ($\phi$) or 180° ($\phi + \pi$) for simulating a logic 1 or 0, respectively.

## S2. Characterizing mode source

Surface plasmon polaritons were excited using the standard mode source in Lumerical Solutions[12] to inject a fundamental guided mode into the plasmonic waveguide. The standard mode source injects a broadband Gaussian pulse signal as shown in Fig. S1(a). Fig. S1(b, c) shows the spectrum of the excitation signal with a center wavelength of 1550 nm and center frequency of 193 THz with a bandwidth of 43 THz. We use this excitation for our simulation throughout the manuscript unless specified.

We additionally ran simulations changing the mode source from a broadband to a narrowband by changing the pulse duration of the Gaussian pulse signal as shown in Fig. S1(d). Fig. S1(e, f) shows the spectrum of the excitation signal with a center wavelength of 1550 nm and center frequency of 193 THz with a bandwidth of 10.7 THz. The simulation results for a 3-input majority logic gate using this narrowband excitation signal is shown in supplementary section S8. The results and the conclusion of the paper remain same for either of the mode sources.

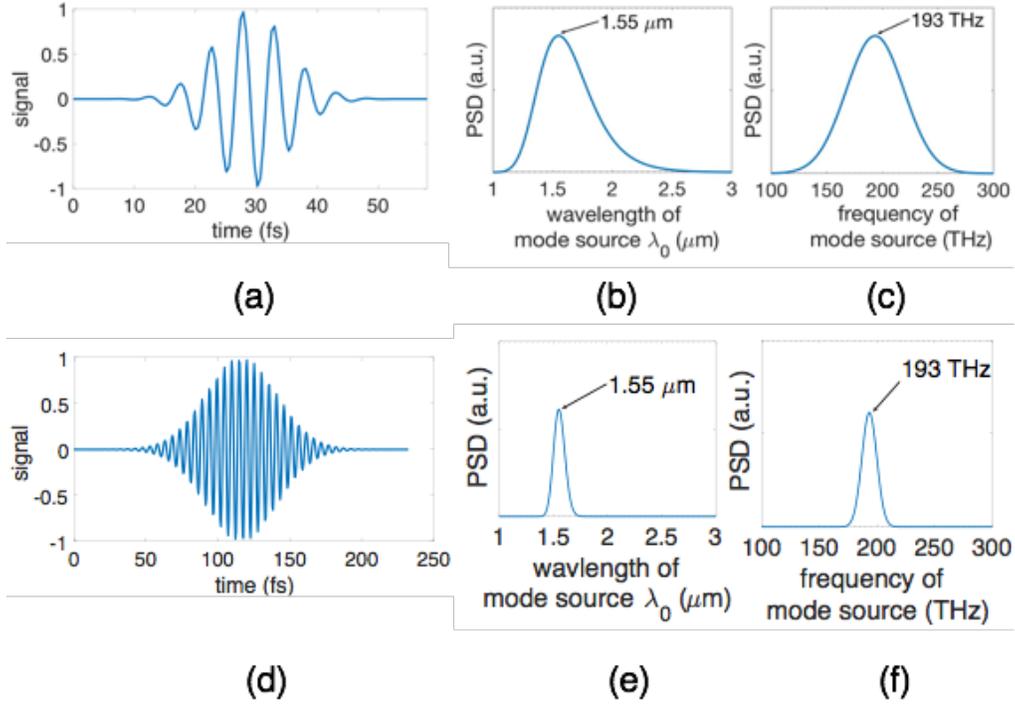

Figure S1. (a) Broadband Gaussian pulse signal injected by the standard mode source in Lumerical simulation. (b, c) Spectrum of the excitation signal with a center wavelength of 1550 nm and center frequency of 193 THz with a bandwidth of 43 THz. (d) Narrowband Gaussian pulse signal injected by the standard mode source in Lumerical simulation. (b, c) Spectrum of the excitation signal with a center wavelength of 1550 nm and center frequency of 193 THz with a bandwidth of 10.7 THz.

**S3. Impact of gap width of the output waveguide on the transmitted power**

Fig. S2(a) shows the increase in the transmitted output power (normalized to the source) with the increase in the gap width of the output waveguide. This results in a considerable reduction of the backflow of power occurring due to reflection from the merging point. The increase in transmission due to wider output gap width can be further explained by resorting to an approach of impedance matching put forward by Cai *et. al* [13]. At wavelengths considerably larger than the size of the structure where the absorption is negligible (above the visible spectrum for silver), one can use the quasi-static approximation treating the waveguides as equivalent transmission lines with some characteristic impedance [13-15]. Following the approach highlighted in ref. [13] for 3-D waveguides, we calculate the direct integrals for the transverse electromagnetic fields to evaluate the effective voltage $v = \int_{-\infty}^{+\infty} E_y dy$ and current $I = \int_{-\infty}^{+\infty} H_z dz$ and hence the equivalent impedance $Z = \frac{v}{I}$. Fig. S2(b) illustrates the reason for the increase in the transmission due to an increase in the impedance of the waveguide (calculated from Lumerical simulations) with the output gap width.

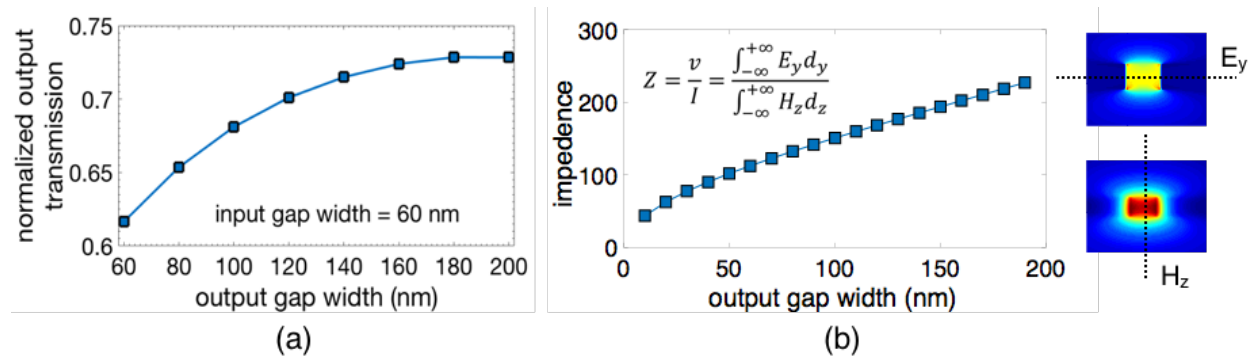

(a)                               (b)

Figure S2. (a) Simulation result showing the increase in the normalized output transmitted power with the increase in the gap width of the output waveguide. (b) Calculated impedance of the waveguide as function of the gap width.

## S4. Time-lapse simulation results for single stage 3-input majority gate

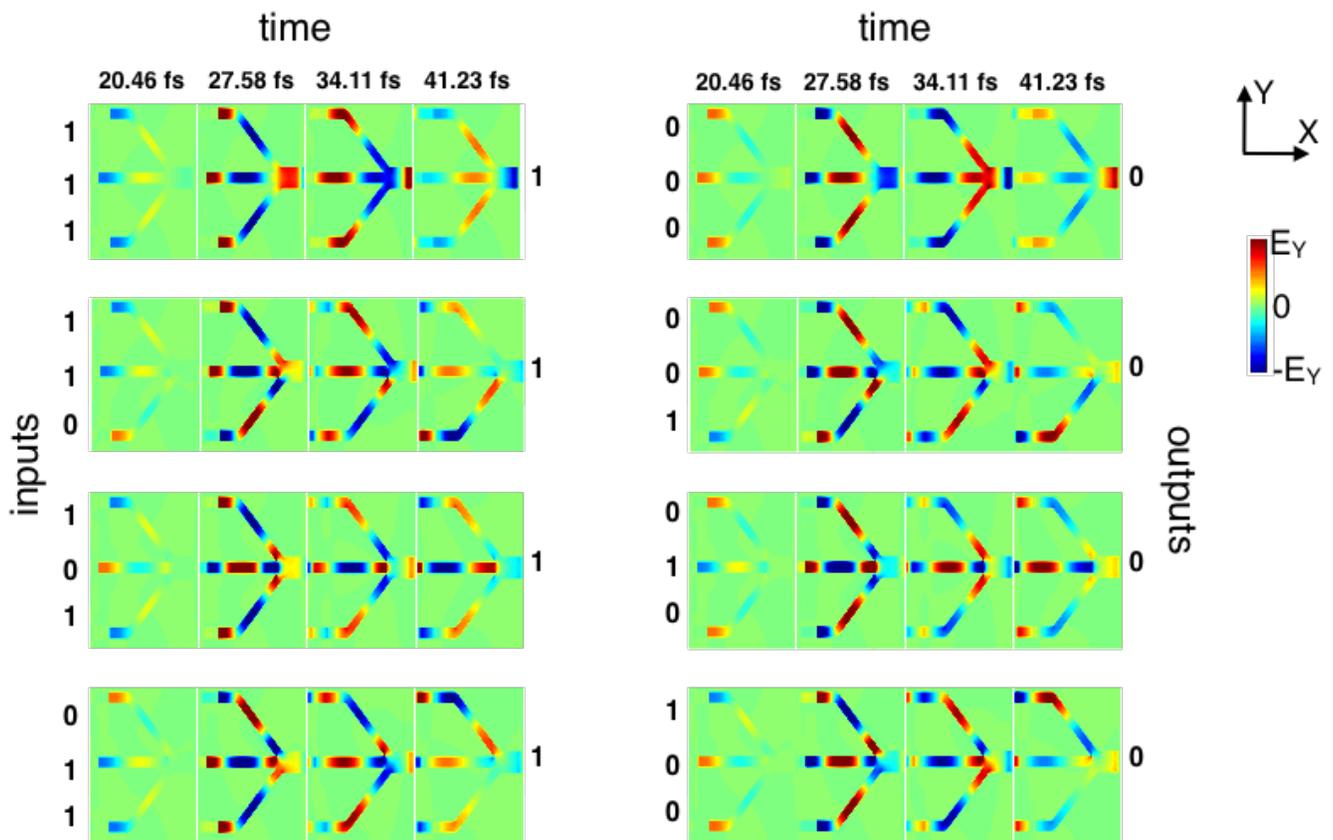

Fig. S3. Time-lapse simulation results in terms of the distribution of the electric field component $E_Y$ in the x-y plane

## S5. Dimensional scaling at each stage

We use the same layout for the 3-input majority gate as seen in Fig. 2(b) of the main paper at each stage but with appropriate dimensional scaling as shown in Fig. 4. Since the pitch follows a scaling of $p$, $3p$ and $9p$ as seen in Fig. 4(a), to keep the same small merging angle of 35° between the waveguides, we also scale the length of the combiner region ($x_2$) accordingly. Note that due to the increase in the length of bends, the path difference ($\delta l_1$, $\delta l_2$ and $\delta l_3$) at each stage has to be considered separately and should be adjusted in the middle arm. For improved transmission via impedance matching, we choose the widths of the waveguides at each stage as $w$ (=60 nm), $2w$ and $3w$ respectively. Note that we resort to a non-aggressive scaling of the widths since beyond $3w$ (=180 nm), the coupled MIM mode tends to split into two separate modes. However, we still gain from the fact that by increasing the widths, we increase the propagation length $L_P$ of the SPP from 5.3 µm in the first stage to 8.46 µm in the second and 10.76 µm in the third. This proves to be beneficial since the overall size of the majority gate increases from one stage to the next.

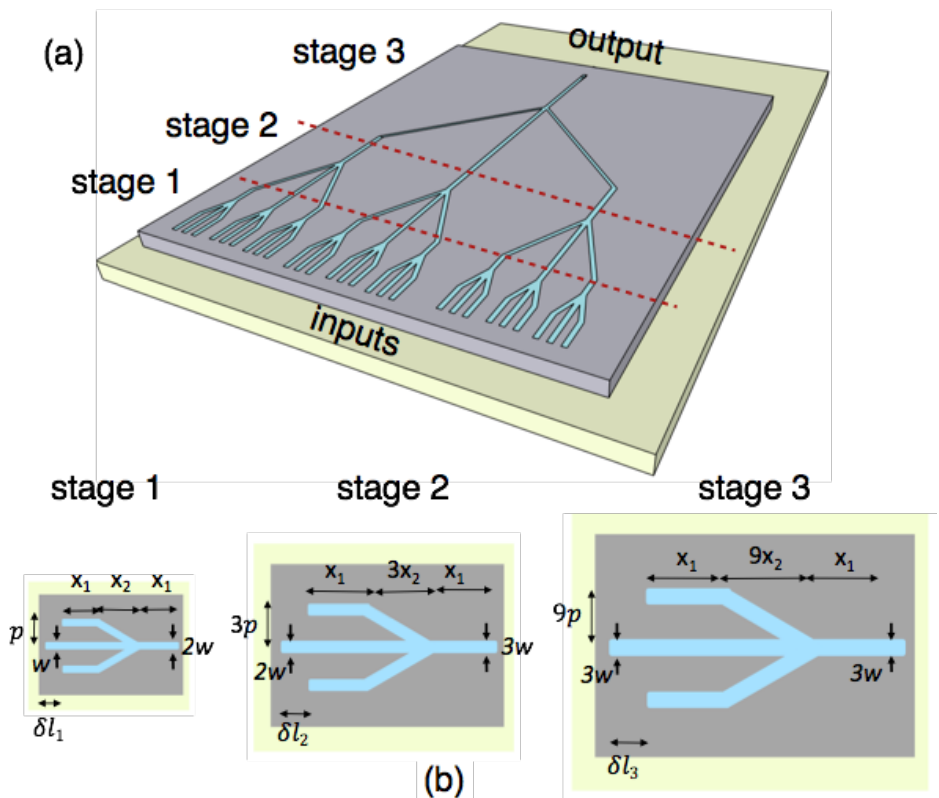

Figure 4. (a, b) Dimensional scaling and layout for cascaded majority gate structure.

## S6. Time-Lapse Simulation Results for 2-Stage Cascaded Majority Logic Gate

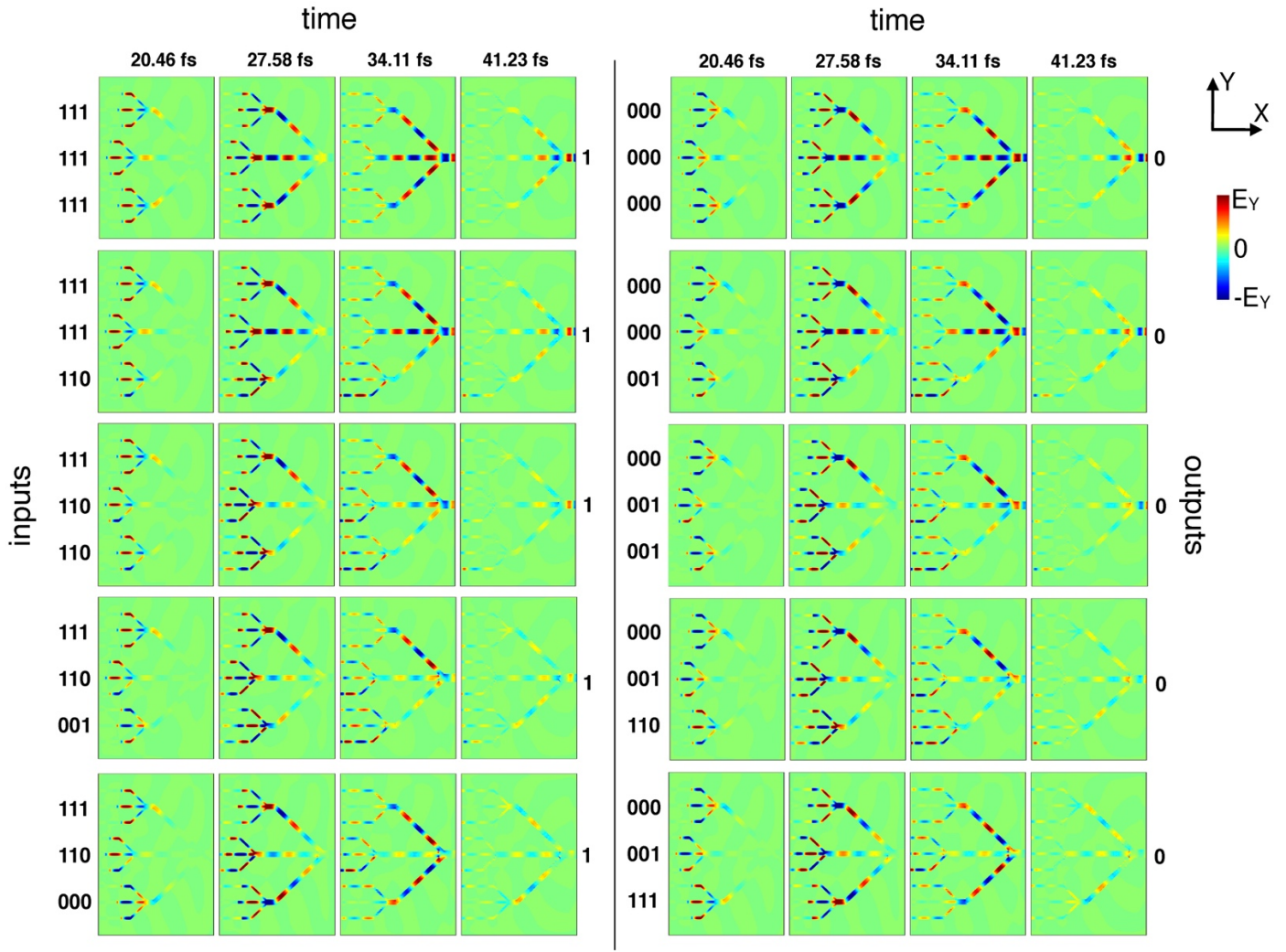

Figure S5. Time-lapse simulation results in terms of the distribution of the electric field component $E_Y$ in the x-y plane

## S7. Time-Lapse Simulation Results for 2-Stage Cascaded Majority Logic Gate with Reference

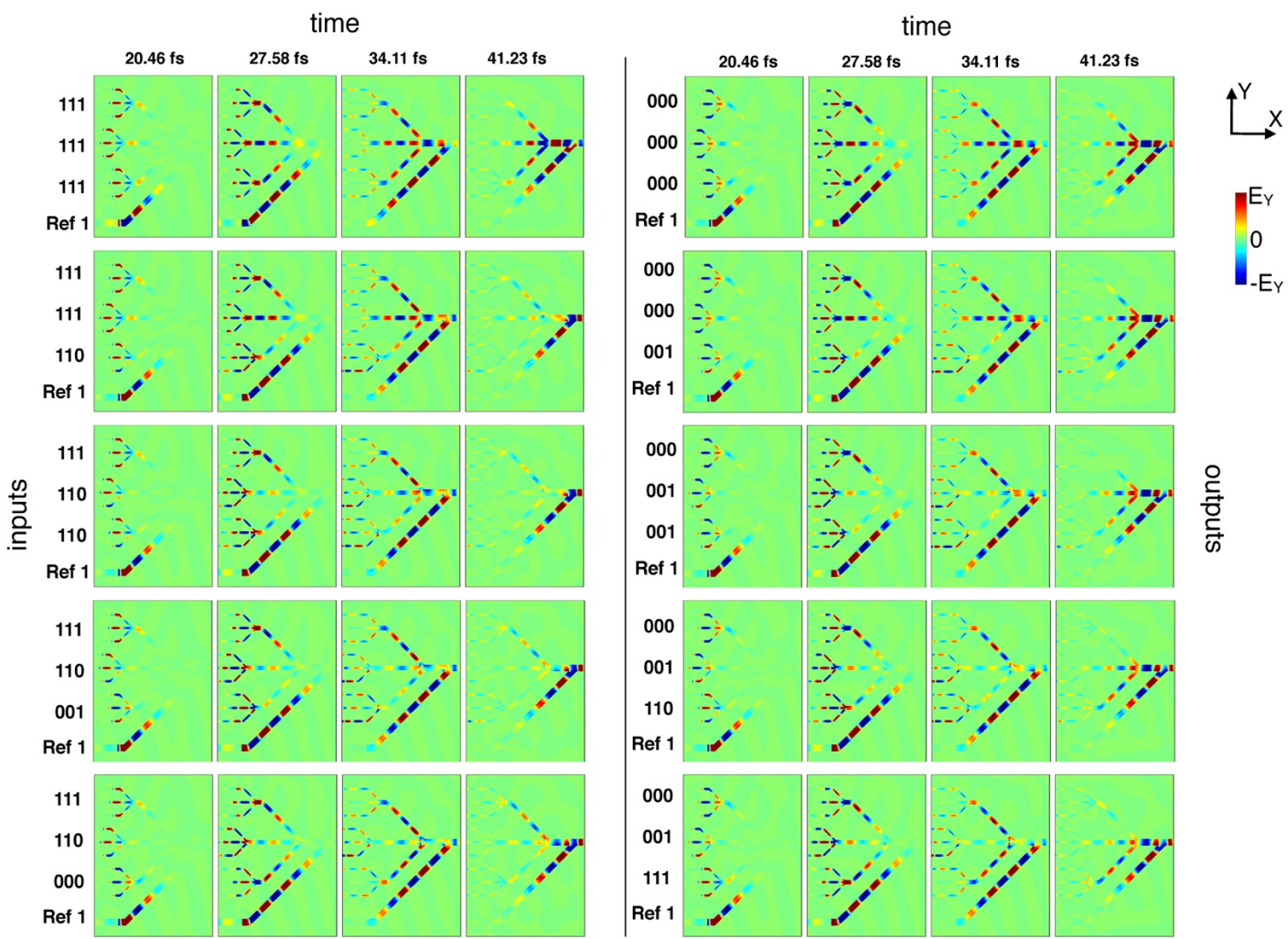

Figure S6. Time-lapse simulation results in terms of the distribution of the electric field component $E_Y$ in the x-y plane

## S8. Simulation results for single stage 3-input majority gate with narrowband excitation

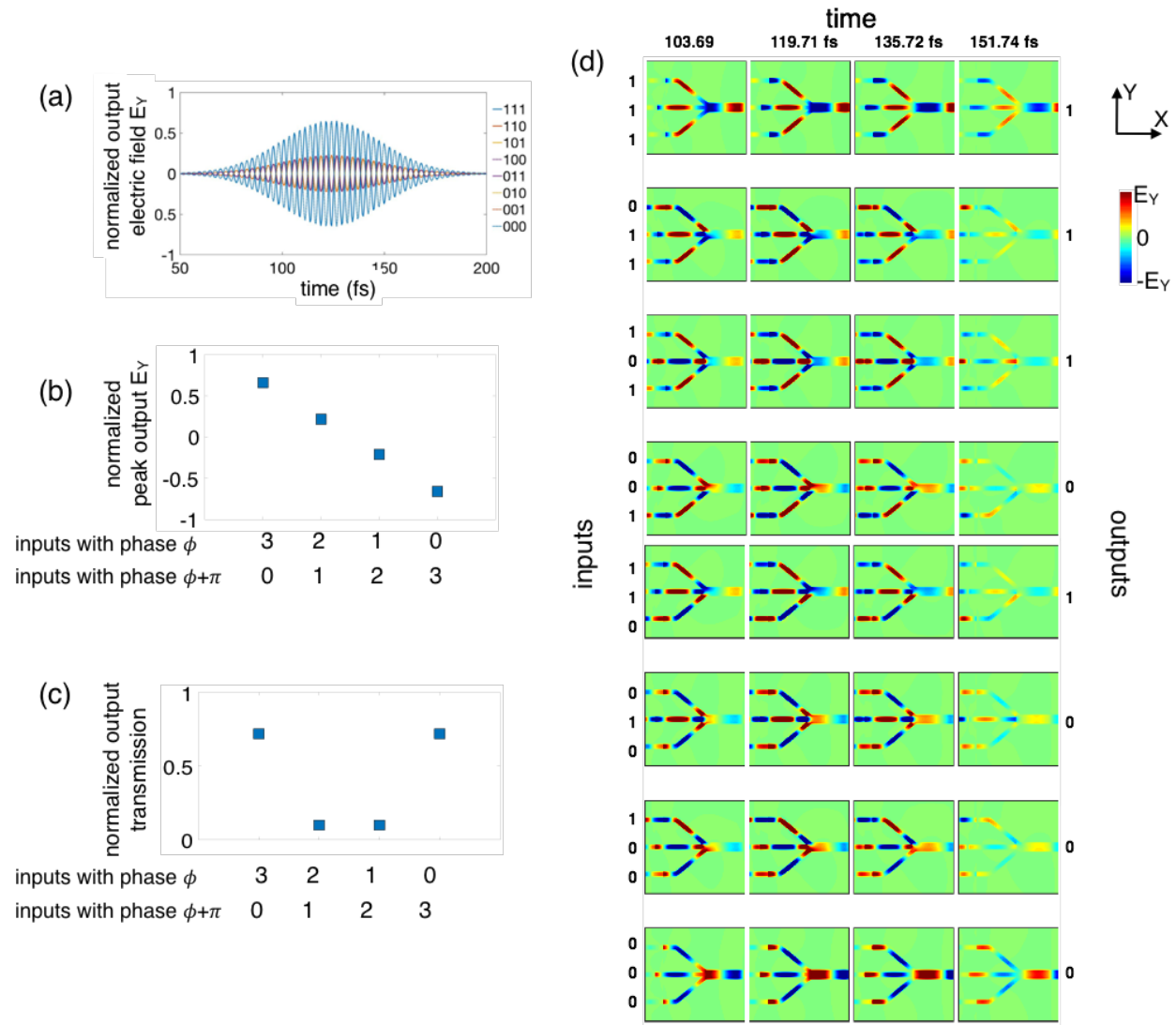

Figure S7. (a) Simulation result for a 3-input plasmonic majority gate for $2^3$ input combinations in terms of the time-domain electric field component $E_Y$ at the output, normalized to the total source electric field and averaged over the cross-section of the output waveguide. A mode source with narrowband excitation is used as mentioned in section S2. (b, c) Calculated peak values of the normalized average electric field component $E_Y$ and normalized transmitted power at the output for different combinations of the input phases. (d) Time-lapse simulation results in terms of the distribution of $E_Y$ in the x-y plane showing the propagation and interference of the SPP waves.

# References

1. Ebbesen, T. W., Genet, C. & Bozhevolnyi, S. I. Surface-plasmon circuitry. *Physics Today* (2008).
2. Maksymov, I. S. & Kivshar, Y. S. Broadband light coupling to dielectric slot waveguides with tapered plasmonic nanoantennas. *Optics letters* **38**, 4853-4856 (2013).
3. Panchenko, E., James, T. D. & Roberts, A. Modified stripe waveguide design for plasmonic input port structures. *Journal of Nanophotonics* **10**, 016019-016019 (2016).
4. Koller, D. *et al.* Organic plasmon-emitting diode. *Nature Photonics* **2**, 684-687 (2008).
5. Neutens, P., Lagae, L., Borghs, G. & Van Dorpe, P. Electrical excitation of confined surface plasmon polaritons in metallic slot waveguides. *Nano letters* **10**, 1429-1432 (2010).
6. Walters, R. J., van Loon, R. V., Brunets, I., Schmitz, J. & Polman, A. A silicon-based electrical source of surface plasmon polaritons. *Nature Materials* **9**, 21-25 (2010).
7. Cazier, N. *et al.* Electrical excitation of waveguided surface plasmons by a light-emitting tunneling optical gap antenna. *Optics express* **24**, 3873-3884 (2016).
8. Uskov, A. V., Khurgin, J. B., Protsenko, I. E., Smetanin, I. V. & Bouhelier, A. Excitation of plasmonic nanoantennas by nonresonant and resonant electron tunnelling. *Nanoscale* **8**, 14573-14579 (2016).
9. Kern, J. *et al.* Electrically driven optical antennas. *Nature Photonics* **9**, 582-586 (2015).
10. Papaioannou, S. *et al.* Active plasmonics in WDM traffic switching applications. *Scientific reports* **2**, 652 (2012).
11. Im, S.-J. *et al.* Plasmonic phase modulator based on novel loss-overcompensated coupling between nanoresonator and waveguide. *Scientific reports* **6** (2016).
12. Lumerical, F. Solutions. *Web source [*https://www/*. lumerical. com/tcad-products/fdtd/]* (2012).
13. Cai, W., Shin, W., Fan, S. & Brongersma, M. L. Elements for Plasmonic Nanocircuits with Three-Dimensional Slot Waveguides. *Advanced materials* **22**, 5120-5124 (2010).
14. Veronis, G. & Fan, S. Bends and splitters in metal-dielectric-metal subwavelength plasmonic waveguides. *Applied Physics Letters* **87**, 131102 (2005).
15. Kocabas, S. E., Veronis, G., Miller, D. A. & Fan, S. Transmission line and equivalent circuit models for plasmonic waveguide components. *IEEE Journal of Selected Topics in Quantum Electronics* **14**, 1462-1472 (2008).